\documentclass[aps,prb,reprint,superscriptaddress,twocolumn]{revtex4-2}

\usepackage{graphicx}
\usepackage[normalem]{ulem}
\usepackage{natbib}
\usepackage{amsmath,array,amssymb}
\usepackage{dcolumn}
\usepackage{bm}
\usepackage{hyperref}
\usepackage{color}
\usepackage{amsmath, amsthm, amssymb, amsfonts}
\usepackage{todonotes}

\graphicspath{{img/}}

\begin{document}
	 \title{Quadratic Zeeman Spectral Diffusion of Thulium Ion Population in a Yttrium Gallium Garnet Crystal}
	%\title{Spectral Population Dynamics of Thulium Ions in a Yttrium Gallium Garnet Crystal}
	%-----------------------------

	\author{Jacob H. Davidson}
	\altaffiliation{These authors contributed equally to this work. Present Address: National Institute of Standards and Technology (NIST), Boulder, Colorado 80305, USA}
	\affiliation{QuTech and Kavli Institute of Nanoscience, Delft University of Technology, 2628 CJ Delft, The Netherlands}
	
		\author{Antariksha Das}
		\altaffiliation{These authors contributed equally to this work.}
	\affiliation{QuTech and Kavli Institute of Nanoscience, Delft University of Technology, 2628 CJ Delft, The Netherlands}
	
	    \author{Nir Alfasi}
	\affiliation{QuTech and Kavli Institute of Nanoscience, Delft University of Technology, 2628 CJ Delft, The Netherlands}
	
		\author{Rufus L. Cone}
	\affiliation{Department of Physics, Montana State University, Bozeman, Montana 59717, USA}
	
		\author{Charles W. Thiel}
	\affiliation{Department of Physics, Montana State University, Bozeman, Montana 59717, USA}

	\author{Wolfgang Tittel}
	\affiliation{QuTech and Kavli Institute of Nanoscience, Delft University of Technology, 2628 CJ Delft, The Netherlands}
	\affiliation{Department of Applied Physics, University of Geneva, 1211 Geneva 4, Switzerland}
	\affiliation{Schaffhausen Institute of Technology - SIT, 1211 Geneva 4, Switzerland}
	\date{\today}
	%-----------------------------	
	
\begin{abstract}
	The creation of well understood structures using spectral hole burning is an important task in the use of technologies based on rare earth ion doped crystals. We apply a series of different techniques to model and improve the frequency dependent population change in the atomic level structure of Thulium Yttrium Gallium Garnet (Tm:YGG). In particular we demonstrate that at zero applied magnetic field, numerical solutions to frequency dependent three-level rate equations show good agreement with spectral hole burning results. This allows predicting spectral structures given a specific hole burning sequence, the underpinning spectroscopic material properties, and the relevant laser parameters. This enables us to largely eliminate power dependent hole broadening through the use of adiabatic hole-burning pulses. Though this system of rate equations shows good agreement at zero field, the addition of a magnetic field results in unexpected spectral diffusion proportional to the induced Tm ion magnetic dipole moment and average magnetic field strength, which, through the quadratic Zeeman effect, dominates the optical spectrum over long time scales. Our results allow optimization of the preparation process for spectral structures in a large variety of rare earth ion doped materials for quantum memories and other applications. 
  
\end{abstract}

\pacs{}
	
\maketitle

\section{Introduction}

   Rare-earth ion doped crystals (REICs) are interesting materials due to their long-lived excited states and their exceptionally long optical coherence times at cryogenic temperature \cite{Babbitt2014, Tittel2010}. In particular, along with the possibility for spectral tailoring of their inhomogeneously broadened 4f$^N$-4f$^N$ transitions, this makes them prime candidates for a number of applications in classical and quantum optics. Examples include laser stabilization, RF spectrum analysis, narrow band spectral filtering, and quantum information storage and processing \cite{Sellin1999,Sellin2001,Colice2004,Berger2016,Tittel2010,Kinos2021}.
    
    Thulium-doped Yttrium Gallium Garnet (Y$_3$Ga$_5$O$_{12}$, Tm:YGG) is one such material. Its $^3$H$_6 \leftrightarrow ^3$H$_4$ transition at 795 nm wavelength features an optical coherence time of more than 1 ms \cite{Askarani2021,Thiel2014PRB,Thiel2014PRL}, which is one of the longest among all studied REICs. In combination with the accessibility of this transition—within the range of commercial diode lasers—this makes it a natural candidate for applications. 
    
    The quality of created features and the resulting consequences for associated applications, are dependent on the spectroscopic properties of the dopant ions and their numerous interactions with other atomic components in their local crystalline environment \cite{Macfarlane1980,THIEL2011}, the details of the optical pumping process, and the spectral and temporal profile of the applied laser pulses\cite{Moerner1988,Allen2012}. Deep understanding of the relation between spectroscopic properties, optical control fields, and spectral diffusion dynamics has resulted in improvements of this process in a number of other rare-earth-doped materials including Tm:YAG, Eu:YSO, and Pr:YSO \cite{Bonarota2010,Jobez2016,Rippe2005}. However, this important connection has thus far not been made for Tm:YGG.

    In this paper we track the evolution of population within the electronic levels of Tm$^{3+}$ ions in YGG (see Fig. \ref{CrystalCell} for simplified level scheme) by semi-continuous monitoring of spectral holes for many sequences of applied spectral hole burning pulses. The characteristic shapes and sizes of these spectral features are matched to a rate equation model that encompasses the ground ($^3$H$_6$), excited ( $^3$H$_4$), and bottleneck ($^3$F$_4$) levels in this material with associated lifetimes and branching ratios. At zero magnetic field we see good agreement between our numerical model and measured results across many different pump sequences of varying duration, power, and spectral shape. With the addition of an external magnetic field the agreement with our numerical model disappears as spectral diffusion from local host spins begins to dominate the shape of all spectral features over long timescales. We characterize the nature of this unexpected behavior and expand our model accordingly by adding a spectral diffusion term to account for a quadratic Zeeman interaction with present noisy magnetic fields \cite{Veissier2016_quadratic,Bottger2006}. 
    
    The letter is structured as follows: In section \ref{Exp Setup} we describe the experimental setup used to collect our measurements. In section \ref{Tm Site Structure} we detail the atomic level structure in Tm:YGG and introduce spectral hole burning, the workhorse of our investigations, to select a known set of atomic population. In section \ref{No field Rates} we introduce and apply a rate equation model which shows good agreement to the measured spectral hole features. In section \ref{adiabatic shaping} we detail the use of adiabatic pulses to shape spectral holes at zero magnetic field with the goal of creating high-resolution features. Section \ref{Mag Noise} shows un-controlled changes to created spectral holes in the presence of magnetic fields and connects these noise effects to the quadratic Zeeman effect. Section \ref{Spec diff} extends this quadratic Zeeman connection to the characterization of spectral diffusion results that differentiates the measured results from those predicted by our model over longer timescales.

\section{Experimental Setup} \label{Exp Setup}
\begin{figure}[h!]
\includegraphics[width=0.5\textwidth]{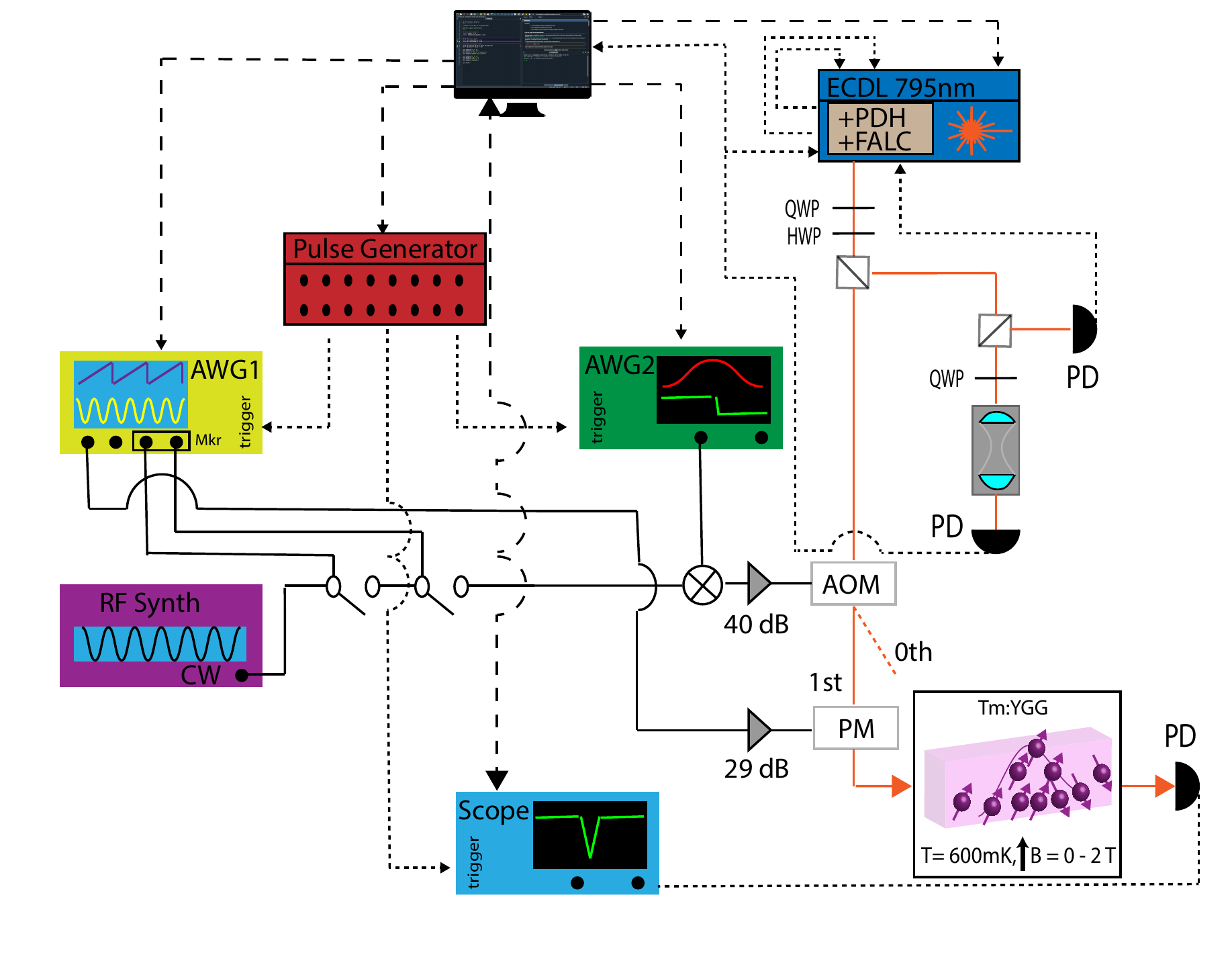}	
\caption {Schematic of the experimental setup.  A PC programs a sequence on a pulse generator (SpinCore Pulseblaster) with nanosecond timing resolution that produces a set of trigger pulses for all devices. Waveforms written to an arbitrary waveform generator (AWG1,Tektronix AWG 70002A) voltage channel, are subsequently amplified (SHF S126 A), and drive an electro optic phase modulator (PM) to generate side bands on the laser light at arbitrary frequencies. The AWG marker channels drive a set of home-built electrical switches which gate the drive signal of an acousto-optic modulator (AOM) to create short pulses from the laser light. This gated AOM signal is mixed with fast arbitrary voltage pulses(AWG2,Tektronix AFG 3102), amplified (Mini-Circuits ZHL-5W-1+), and sent to an acousto-optic modulator (AOM Brimrose 400MHz) to synthesize controllable-amplitude laser pulses with rise times as short as 2.5ns. We use a single light source (Toptica DL Pro 795nm) for different tasks (optical pumping, pulse generation, etc). The laser frequency is set via a wavemeter (Bristol 871) and locked to a thermally and acoustically isolated high finesse optical cavity (Stable Laser Systems) via the Pound-Drever-Hall method and a fast feedback loop acting on the laser current and piezo voltage (Toptica PDH and FALC Modules). Transmitted signals from the crystal are directed to a variable-bandwidth photo detector (NewFocus 2051) and displayed on an oscilloscope (Lecroy Waverunner 8100A) configured by the PC and synchronized with the experimental sequence by a trigger signal.}
\label{Pumping Setup}

\end{figure}

To measure spectral holes, from population storage in various atomic levels, over different timescales in Tm:YGG we use the setup detailed in Fig. \ref{Pumping Setup}. A CW diode laser tuned to the ion transition frequency at 795.325nm \cite{Thiel2014} is locked to a reference cavity resulting in a linewidth of roughly 5kHz \cite{Black2001}. To craft short pulses of high extinction ratio, its continuous wave emission is directed to a free space AOM. The sinusoidal driving signal of the AOM is mixed with a signal modulated by an arbitrary function generator, which allows programmable control of the transmitted pulse amplitude for the first order light. 
	
	After the amplitude control, the pulsed light is directed to a fiber coupled phase modulator driven using arbitrary waveforms for serrodyne frequency shifting and more complex chirped pulse shapes as detailed in section \ref{adiabatic shaping}. The optical signals are then sent through a polarization controller to a 1\% Tm:YGG crystal grown by Scientific Materials Corp. and housed in a pulse tube cooled cryostat at 500-700mK. A superconducting solenoid centered on the crystal applies a homogeneous magnetic field from 0-2T(using about 1mA/mT of current) along the crystal's $<$111$>$ axis.  Signals transmitted through the crystal are directed to a fiber-coupled photo diode and recorded for subsequent analysis

	Experimental control is handled on a number of different time scales via custom Python scripts that ensures signals are created at the correct moment \cite{DavidsonControl2021}. For sequencing on timescales of longer than a second, the built-in Python timing functions are used to adjust the experiment. On all timescale shorter than seconds, timing is handled by pre-programming a pulse generator that produces correctly timed trigger signals for the various devices. Waveforms for the arbitrary voltage signals are generated by custom scripts and uploaded to the respective devices for arbitrary control of instantaneous pulse frequencies and amplitudes.

\section{Tm:YGG Site and Level Structure} \label{Tm Site Structure}

    Garnet crystals such as YGG have cubic crystal structure with O$^{10}_{h}$ space group symmetry, which yields six Tm$^{3+}$ ion substitution sites, each with D$_2$ point group symmetry\cite{Thiel2014PRB,Dillon1961, Menzer1929, Sun2005}. The magnetic and optical behavior of Tm$^{3+}$ ions in each of these sites is identical but the crystal structure leads to effective in-equivalence between the sites due to six different orientations that the ion and its entire local environment can take within the lattice \cite{Davidson2021}.  However, for a few specific directions relative to the crystalline axes, ions at in-equivalent sites can be cast into classes that share the same projections of applied electromagnetic fields ($\boldsymbol{\vec{E}},\boldsymbol{\vec{B}}$) onto their local site axes.
    
\begin{figure}[h!]
\begin{centering}
\includegraphics[width=0.5\textwidth]{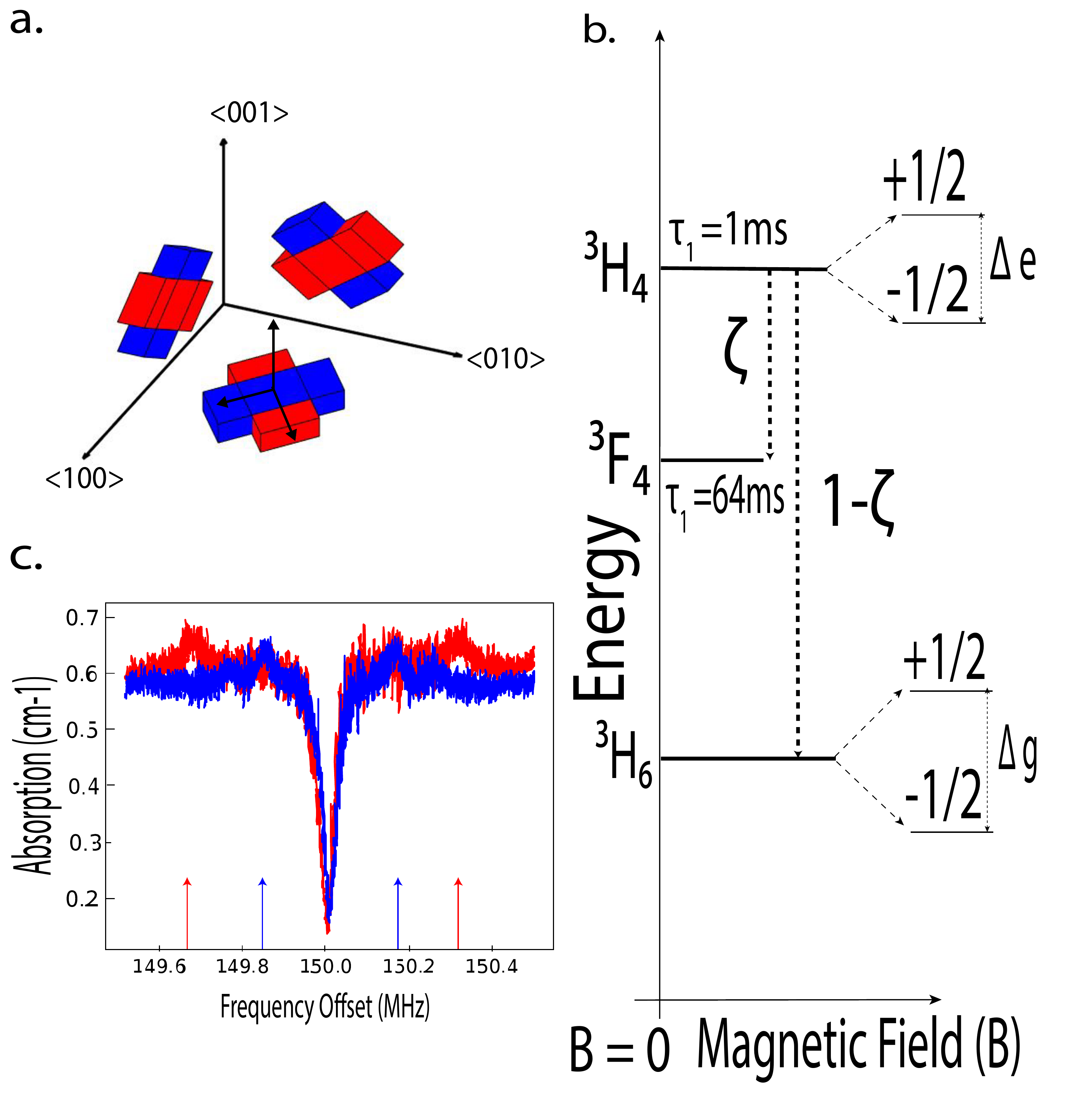}	
\caption {\textbf{a.} Depiction of the six ion substitution sites relative to the cubic crystal cell. For a magnetic field along the crystalline $<$111$>$ axis the sites are cast into two classes featuring different field projection, shown in red and blue. Small black arrows indicate the local site X,Y,Z, axes for the red site in the $<$100$>$ $<$010$>$ plane. \textbf{b.} Energy level structure of the Tm ion transition of interest with previously measured lifetimes ($\tau$), branching ratios ($\zeta$), and ground and excited splittings ($\Delta g$, $\Delta e$).  \textbf{c.} A pair of hole-burning spectra with the main spectral hole pictured in the center. Each hole, shown in an associated color to part \textbf{a.}, is burned with a polarization that selects one of the two classes of ions and produces anti-holes with different splittings (indicated by the colored arrows). Additional modulations of the optical depth outside the hole originate from a slight orientation offset during crystal cutting and polishing.}
\label{CrystalCell}
\end{centering}
\end{figure}

    Given the orientation of our magnetic field, parallel to the crystalline $\boldsymbol{\vec{B}}$ $|| <$111$>$ axis, we cast the ions into two different classes as depicted in Fig. \ref{CrystalCell} a. The ions in each separate class experience different magnetic field projections onto the axes in their local frame. One class features magnetic field projections along the ion's local X and Z axes (blue), and the other projections along the local Y and Z axes (red). This difference becomes evident from a simple spectral hole burning experiment, which we use to introduce the remaining results of the paper.
    
    In spectral hole burning, a long optical pump pulse excites atomic population in a narrow spectral window within an inhomogeneously broadened absorption line. Excited ions subsequently decay — either back into the original state, or into another energy level that often belongs to the electronic ground state manifold. Scanning a weak laser beam over a spectral interval centered on the frequency of the original pump pulse reveals sections of decreased and increased absorption — so-called spectral holes and anti-holes. Spectral holes occur at frequencies of reduced ground-state population, i.e. with offset $\Delta = 0$ (for the central hole), $\Delta = D_e$ (for the side hole), and anti-holes can be observed whenever the ground-state population is increased, which happens at $\Delta = D_g$ and $\Delta = D_g \pm D_e$. Here, $D_g$ and $D_e$ are ground and excited state splittings.  Consult Ref.\cite{Lauro2009, Louchet2007,Moerner1988}  for more details on spectral hole burning.

     In our case, the ground and excited state splittings depend on the magnitude and direction of the applied magnetic field — which vary for each class of Tm ions \cite{Davidson2021}. At 7.5 mT we recorded a pair of hole burning spectra, shown in Figure \ref{CrystalCell} c, for orthogonal pumping polarizations. This allowed us to selectively address ions in either of the two classes. We found two sets of anti-holes, each of which split according to the different field projections experienced by the two classes of Tm$^{3+}$ ions. This ability to select out a single class of ions becomes important in section \ref{Spec diff} as spectral diffusion depends on the magnetic projection on each specific class of ions.

\section{Modeling Results Using Three-Level Rate Equations}	\label{No field Rates}

    To further understand the effects of a given hole burning process on our REIC ensemble we turn to solutions of the Maxwell-Bloch equations that describe the interaction of light with one or many atomic systems. These differential equations can be quite difficult to solve, given the complexity that there does not exist a single fixed Rabi frequency to drive all ions \cite{Sun2000}.  In the limit of excitation pulse lengths much shorter or much longer than $T_2$ the rate equation approach has been shown to be an effective model \cite{Thiel_ISD2014}. Thus for the case of Tm:YGG, due to the long $T_2$ and low optical depth we reduce the Maxwell-Bloch equations to a set of rate equations that describe the conserved total atomic population and how it flows through the different available levels as a function of time \cite{Allen2012}.  
    
    Note that in the case of narrow band excitation of an inhomogeneously broadened transition where only a certain portion of atoms are driven, frequency dependence must be added. Following \cite{Allen2012} and \cite{Bonarota2010,Linget2015} we describe the dynamics of our atomic ensemble with equations \ref{eq1}-\ref{eq3}.
	
\begin{align} \label{eq1}
\frac{\partial n_g(t)}{\partial t} &= R(\Delta)(n_e - n_g)+\frac{1-\zeta}{T_{e}}n_e + \frac{1}{T_{b}}n_b\\
\label{eq2}
\frac{\partial n_e(t)}{\partial t} &= R(\Delta)(n_g - n_e)-\frac{1}{T_{e}}n_e\\
\label{eq3}
\frac{\partial n_b(t)}{\partial t} &= \frac{\zeta}{T_{e}}n_e  - \frac{1}{T_{b}}n_b
\end{align} 
	
	This system of coupled differential equations describes the relative change of atomic population, $n_{(g,e,b)}(t,\Delta)$ in three ion levels as a function of time, and frequency offset, $\Delta$.  The branching ratio $\zeta$ determines how much population decays from the excited state $|e\rangle$ through the bottleneck state $|b\rangle$ before reaching the ground state $|g\rangle$ with respective level lifetimes $T_e$ and $T_b$. To drive the system, all frequency dependence in contained in the excitation rate $R(\Delta)$ which characterizes the applied pulses of light used for hole burning. For Tm:YGG, the level lifetimes and branching ratios are well known at zero magnetic field \cite{Thiel2014} leaving only one unknown, the frequency dependent optical pumping rate which has to be matched with measured data. 
	
	This pumping rate $R(\Delta)$ is a function of both the individual ion properties and the applied sequence of laser excitation, or burning, pulses. Again from \cite{Bonarota2010}, 
\begin{align}
	R(\Delta) \propto \gamma \otimes |\mathbb{F}(\Delta)_{Opt}|^2
\end{align}
    is a convolution of the two factors, the homogeneous linewidth $\gamma$ of the ionic transition and the positive frequency component of the Fourier spectrum of the applied pulse sequence $\mathbb{F}(\Delta)_{Opt}$. For the case of Tm:YGG the homogeneous linewidth of the ions is expected to be narrower than 600Hz \cite{Thiel2014}.  For any $\mathbb{F}(\Delta)_{Opt}$ with features greater than the homogeneous linewidth of the ions this convolution is dominated by the Fourier spectrum of the pumping pulse, which is determined by laser and sequence characteristics such as pump power, pulse duration, pulse spacing, and spectral line-shape.

\begin{figure}[t!]
\begin{centering}
\includegraphics[width=0.5\textwidth]{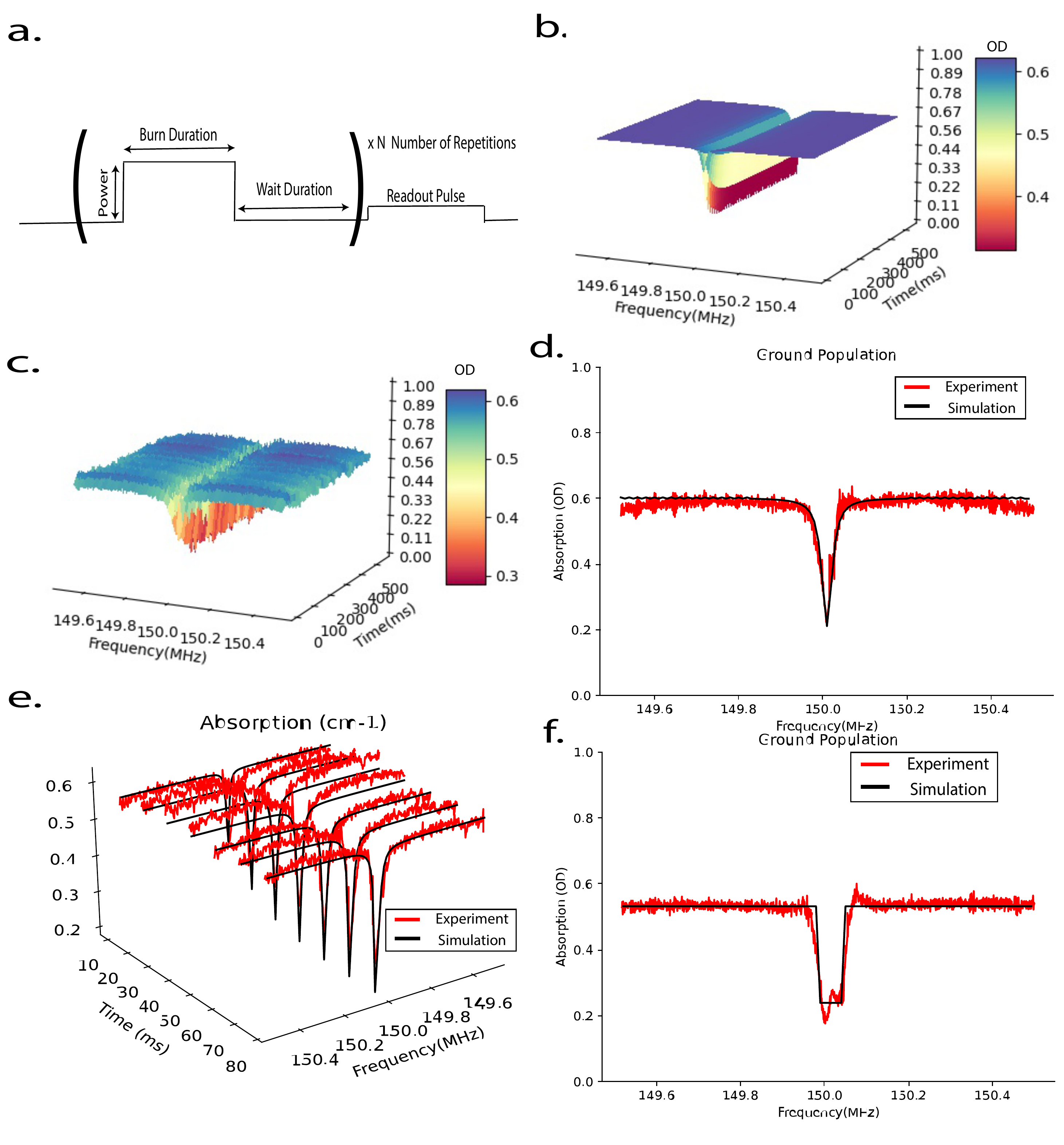}	
\caption {\textbf{a.} Simple hole-burning sequence used throughout the paper. \textbf{b.} Evolution of a spectral hole simulated using Eqs. \ref{eq1}-\ref{eq3}. The hole is detuned by 150MHz from the peak of the Tm absorption line and develops slowly in time. \textbf{c.}  Measured hole evolution. The hole-burning sequence is as in b. The hole deepens as more burn pulses are applied. \textbf{d.} The final hole of both methods( experiment and simulations) after 50 cycles of alternating 1ms long pump pulses and 10ms waiting periods. \textbf{e.} Overlaid slices at many times throughout the burning process.\textbf{f.} Square shaped spectral hole generated using hyperbolic secant burn pulses. Modulations at the bottom of the hole are due to imperfections in the burning pulses, and the square line shows comparison to modeled results from a perfectly square $R(\Delta)$ function of matching bandwidth.}
\label{HoleSurface}
\end{centering}
\end{figure}

	We perform many different hole burning experiments with differently timed burning sequences of various powers, durations, and waiting periods. See figure \ref{HoleSurface} a. With these parameters set, and the known lifetimes and branching ratios of Tm:YGG, we numerically solve the rate equations above for all times throughout the burning sequence after setting the initial conditions before pumping to $n_{g}=1$(i.e. after initializing all population in the ground state). An example is shown in Figure \ref{HoleSurface} b, showing the predicted development of a spectral hole over time for some chosen $R(\Delta)$ parameter. The measured evolution of the spectral hole after the same series of pumping pulses and waiting times is shown in Figure \ref{HoleSurface} c where the spectral hole is read out after the every burn pulse to capture the evolution.
%%% Add different surfaces for part b/c to make hole shape less trivial looking or a different angle. 
    
    As the optical depth of a measured spectral hole is proportional to the ground state atomic population left to absorb, we normalize the remaining population using the optical depth before any pumping has occurred and match the results generated by experiment and by numerical simulation for a particular burning sequence \cite{Bonarota2010,Allen2012}. Note that since we do not have an analytical function which describes the solution to the system of rate equations, the matching is not a fit that outputs a mean standard error. However, as a "goodness of fit" metric, we calculate the point-wise average difference between the measured and simulated surfaces. Solutions and measurements are shown for a few different times in Figures \ref{HoleSurface} d,e to demonstrating close agreement for matched parameters. More examples of different burning sequences, a discussion of the resulting holes with respect to the level lifetimes, and a note on the power dependence of the laser line-shape are included in the supplementary information.
    
     We consider many different $R(\Delta)$ functions to validate our model, only one of which creates the closest match with all the measured results. This allows us to uniquely identify the $R(\Delta)$ parameter as a property of our laser source and its interaction rate with the ions. 
	
	All burning sequences assume a source with a Lorentzian shaped laser line of  $\nu = 5$ kHz linewidth and a excitation rate amplitude $a$, tunable between approximately 300 Hz and 2.5 kHz, via AOM drive power. This yields an $R(\Delta)$ function of the form
\begin{align*}
	R(\Delta) &= a \frac{\nu^2}{\nu^2 +(\Delta -\Delta_{o})^2}  \\
	\quad \text{for}\quad  a &\in [0.3,2.5] \text{kHz},\\
	\nu &= 0.005 \pm 0.002 \text{MHz},\\
	\Delta_{o} &= 150 \pm 0.01\text{MHz}
\end{align*}
    where $\Delta_o$ is the detuning from the thulium absorption line center.
    
	Thus our model for population change driven by this known optical spectrum allows us to predict the size and shape of any spectral feature resulting from a given burning sequence in this material at zero external magnetic field. This strategy extends to other materials as long as the relevant branching ratios and lifetimes are known.

\subsection{Adiabatic Pulse Shaping}	 \label{adiabatic shaping}

    To confirm our measurements, and exclude spectral hole broadening mechanisms from the optical pumping process, we sought to change the spectral shape of our excitation light and create controllable changes to the width and shape of the resulting spectral holes. We use adiabatic pulse shaping to isolate the shape and width of created spectral features for further optimization while continuing to match to solutions of our predictive model.
    
    An important method for controlling the shape and width of a desired spectral feature stems from an early goal of NMR to design a pulse which creates highly efficient yet highly selective population inversion \cite{Silver1985}. The resulting coherent population inversion over a narrow frequency bandwidth is described by a unique analytical solution to the Maxwell-Bloch equations \cite{Hioe1984}.  This method is known as a hyperbolic secant pulse and relies upon slow adiabatic tuning of the amplitude and phase of the driving pulse. This results in the elimination of unwanted frequency components around the desired bandwidth, and the creation of a pulse with a particularly square spectrum \cite{Allen2012,Tian2011}. For REICs, hyperbolic secant pulses were first considered for quantum computing in order to create high fidelity $\pi$ pulses for narrow bandwidth ensembles of ions \cite{Roos2004, Rippe2005}.

    % How we craft them in practice with multiple devices and synchronization description.
    
    The original Maxwell-Bloch solutions call for simultaneous slow modulation of the amplitude and phase, each with a different shape. This is equivalent to modulation of the pulse amplitude and frequency shifting during the duration of the pulse \cite{Silver1985}. From a hardware standpoint the problem becomes creating fast electronic control of the amplitude $A(t)$, via AOM voltage, and instantaneous frequency $f(t)$, via phase modulator drive, as in \cite{Jobez2016}. To account for the separate elements responsible for each modulation, a variable delay was added to the amplitude modulation to ensure that the correct portion of each pulse is temporally aligned with the corresponding frequency shift.

    This method of creating spectral features is in contrast to the common method of linearly swept serrodyne modulation \cite{Tian2011, Johnson1988,Saglamyurek2014}. %To show the benefit we picture example Fourier transforms of pulses crafted by both methods in figure \ref{SecHypeSurface} f. 
    For broadband features both methods are fairly similar, but for creating narrow-band spectral features on the order of the pump laser linewidth, the spectrum of the adiabatic pulse shows much steeper rising and falling edges than that of linearly shifted pulses.

    To show the clear difference in spectral hole shape using this method, we carry out the same hole burning experiment as above with the key difference being that each burning pulse is subject to adiabatic amplitude and frequency modulation. The simulated and measured results of these hole burning experiments are compared in Figures \ref{HoleSurface} f. The hole shape clearly changes corresponding to the altered---squarer---excitation spectrum $R(\Delta)$.  The shape of the resulting hole spectrum is controllable in terms of width by altering the frequency range and speed of the hyperbolic secant pulse modulations. Though the lower limit of hole width is still related to original laser linewidth, adiabatic modulation suppresses Lorentzian wings of the spectrum, and will allow for added power in the burning process. 
    
\section{Magnetic Noise} \label{Mag Noise}
    
    With the ability to control the frequency chirp range and the amplitude and duration of each burning pulse, creation of many possible spectral features, at zero magnetic field, is well understood from the model in Eqs. \ref{eq1}-\ref{eq3}. However, the lifetime of the $^3$F$_4$ the bottleneck level in Tm:YGG creates an upper limit for burning deep yet sharp holes with a reasonable laser power, see Supplementary Material. To improve the depth and lifetime of the resulting features, we now add a magnetic field to create a longer-lived bottleneck state; a ground state spin level with a lifetime of seconds\cite{Davidson2021,thiel2014Tm:YGG}. 
    
     It is worth noting at this point that $R(\Delta)$ is the main term that determines frequency dependence in the model. Thus, with applied magnetic field an identical bandwidth of initial population is excited by the same pulse sequences carried out at zero field. Ideally, only the duration of the subsequent decay path from the excited state can change as magnetic fields are applied  leading to deeper longer lived spectral holes. However, it will be seen in the following section(s) that though we can initially create spectral features of a shape matching the zero field case, quadratic Zeeman coupling between magnetic field noise and the local spin environment of the Tm ions creates significant changes to these features over long time scales. 
    
%%% What happens when we add magnetic field 
    With the addition of a magnetic field the ground and excited states of Tm:YGG each split into spin states with a spin-Hamiltonian given by \cite{Davidson2021}. The created ground state spin level has a much longer lifetime than the $^3$F$_4$ state previously used as the bottleneck level, prolonging eventual decay to the ground state \cite{Thiel_2014}.
    
    A significant portion of electronic population from the $^3$F$_4$ level decays to the additional spin level before reaching the original ground state, creating an equivalent three level system that should be governed by similar rate equations, albeit with extended bottleneck level lifetime \cite{Bonarota2010}. In addition, to ensure that population accumulates in this level, waiting times between burn pulses must be on the order of the $^3$F$_4$ state lifetime to prevent population storage in the $^3$F$_4$ state from where it can decay to the original ground state much more quickly. Measuring the created spectral holes over these longer burning and waiting sequences ---without significance without applied field--- we find important differences compared to the modeling. To assess these differences, spectra in this section and in section \ref{Spec diff} are taken post burn sequence on a longer timescale than those presented in the previous sections. More specifically, we measure a created hole periodically every 11 seconds for minutes. Each hole is fitted to a Lorentzian to determine the center, depth, and width of the spectral feature. 
  
\begin{figure}[h!]
\begin{centering}
\includegraphics[width=0.5\textwidth]{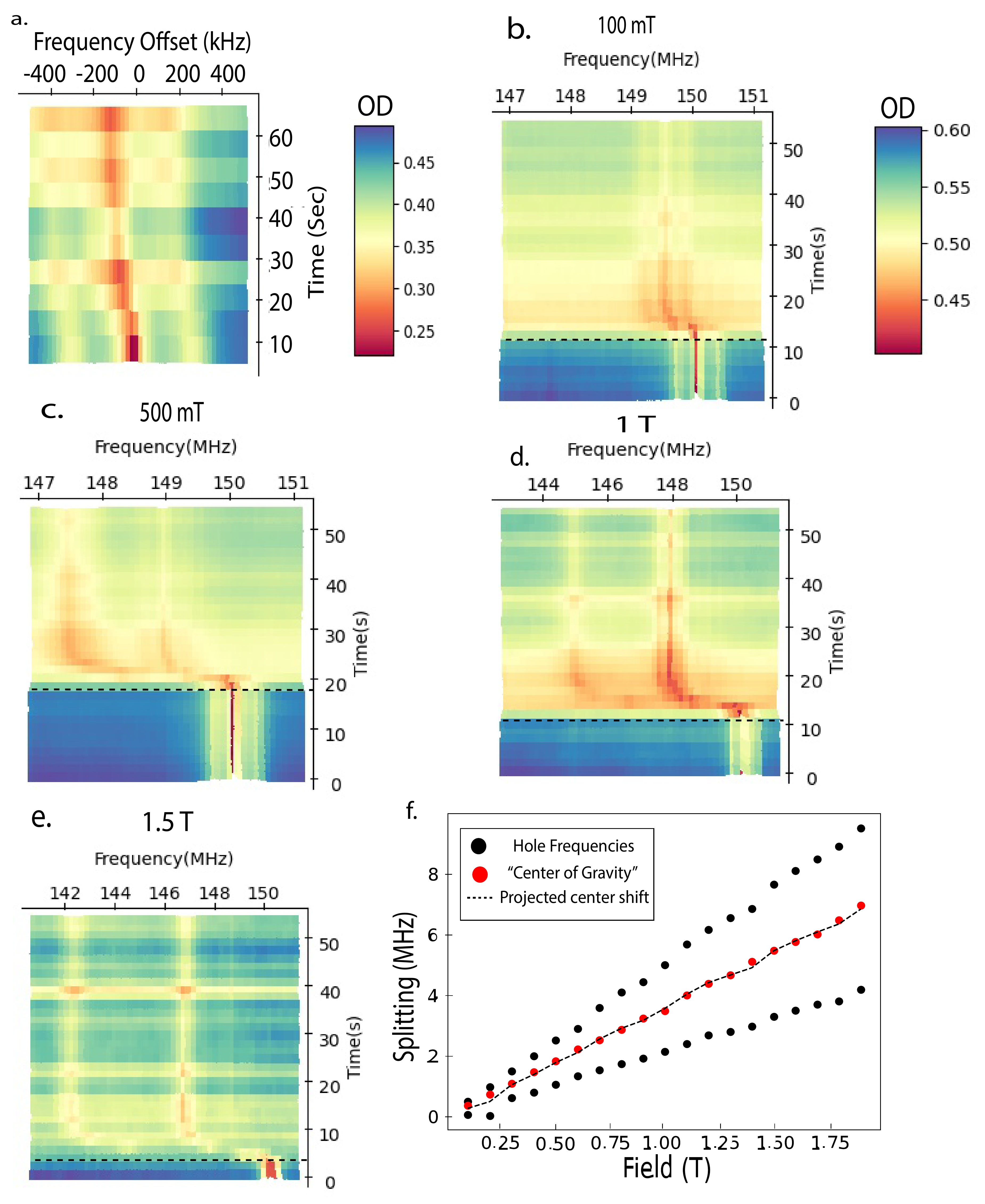}
\caption {\textbf{a.} Spectral hole crafted at 1.8 T, and read out repetitively over the course of one minute. \textbf{b.-e.} Shifting and splitting of the hole features for a fixed change of 1.6mT at initial magnetic fields from 0.1,0.5,1,1.5 T. The dashed lines show the moment when the small change in field was made. \textbf{f.} Hole center frequency shift and splitting into separate hole features after a fixed change of 1.6mT at initial magnetic fields from 0.1-2 T. The dashed line is the center shift projected by the hyperfine tensors for this magnitude of field change.  The magnitude of the center shift and the splitting both grow as expected from \cite{Davidson2021} for higher absolute fields.}
\label{MNoise}
\end{centering}
\end{figure}

    As the magnetic field strength is increased we observe a clear impact of magnetic field noise on the width and center frequency of the spectral holes. This is evident from figure \ref{MNoise} a.
    Here the hole is initially centered at the programmed frequency, but the shape and center proceed to change in an uncontrolled manner due to fluctuations of the applied magnetic field. 
 
    To confirm the origin of the measured effect we create a narrow spectral hole at a fixed magnetic field. A short time after burning has completed, we alter the field by a fixed amount of 1.6 mT, the smallest increment possible with our experimental setup, and measure the subsequent changes to the hole. We expect a splitting and shifting \cite{Sun2012} from the linear and quadratic Zeeman terms of the ion spin-Hamiltonian as ions change optical transition frequency in the altered field. The results of these measurements are depicted in Figs. \ref{MNoise} b-f. We find that the splitting and shifting of the spectral hole increases notably as the value of the initial magnetic field was changed from 0.1 to 1.5 T, for constant field changes of 1.6mT. 

    The measured values of the center shift, and splitting of the pair of resulting holes are in good agreement with expectations from the quadratic and linear components of the hyperfine tensor for these ions \cite{Davidson2021}. The supplementary material contains details for calculating the expected shifting from the hyperfine tensors and the applied magnetic field change. From these measurements, we estimate the field fluctuations in our system to be a factor of 10 smaller than our applied 1.6 mT field change, leading to the shifting of the hole center. The implication is that at high magnetic fields, noise on the order of $\mu$T causes many kilohertz of change in the center hole frequency. This result will prove critical for understanding and creating narrow spectral features at high magnetic fields. 

\section{Spectral diffusion} \label{Spec diff}

    In addition to the central frequency shift, all measured holes are broadened compared to expectations from the burning parameters modeled and tested at zero field, a telltale sign of spectral diffusion \cite{THIEL2011}. 
    
%%% it creates spectral diffusion, what is that?
    Magnetic field dependent spectral diffusion of spins is well understood in the field of rare earths, although it has more often been considered for the case of Kramers, rather than non-Kramers ions such as Tm \cite{Bottger2006}. The problem is often cast in terms of interactions in which nearby spins swapping energy with excited members of the ensemble lead to a general loss of coherence and a spreading, or broadening, of spectral features in time \cite{Klauder1962,Mims1968}. Both magnetic field and sample temperature play important roles as they both mediate how much energy is being interchanged directly through spin-spin interaction or through coupling to host phonons\cite{Veissier2016}. On short to medium timescales the characteristic rate and magnitude of these processes are often measured via two-pulse and three-pulse photon echo sequences \cite{Bottger2006}. For longer timescales spectral hole burning can accomplish a similar task by mapping the change of the hole shape and size as a function of time, magnetic field, and temperature \cite{THIEL2011}.

% How do we measure this process. 
    %%%% What do we measure and why
    
    By adjusting the polarization of the pump light, we first isolate ions with a particular spin splitting at a certain magnetic field, as discussed in section \ref{Tm Site Structure}. Again, we perform spectral hole burning measurements where burning pulses and waiting periods are timed to guarantee population arrives in the desired spin level. Once hole burning is complete, the hole is read out after varying waiting times using, as above, a weak probe in order to leave the hole unperturbed.  We determine (i.e. fit) the widths of all holes for a series of different magnetic field amplitudes. Width change as a function of time can be described by \cite{Bottger2006}
\begin{align} \label{sudden Jump}
\Gamma_{hole} \propto \Gamma_{o}+\frac{1}{2}\Gamma_{SD}(1- e^{-R_st_{delay}}).
\end{align}
    Here, $\Gamma_{o}$ is some initial hole width given by $R(\Delta)$ of the burning process, which broadens in time. The spectral diffusion approaches a maximum value of $\Gamma_{SD}$ at a characteristic rate given by $R_s$. The functional form of the magnetic field and temperature dependence for $\Gamma_{SD}$ and $R_s$ can be used to link spectral diffusion to a specific broadening mechanism \cite{Bottger2006,Macfarlane1993,Lutz2016}.
%%% What do the results say
\begin{figure}[h!]
\begin{centering}
\includegraphics[width=0.5\textwidth]{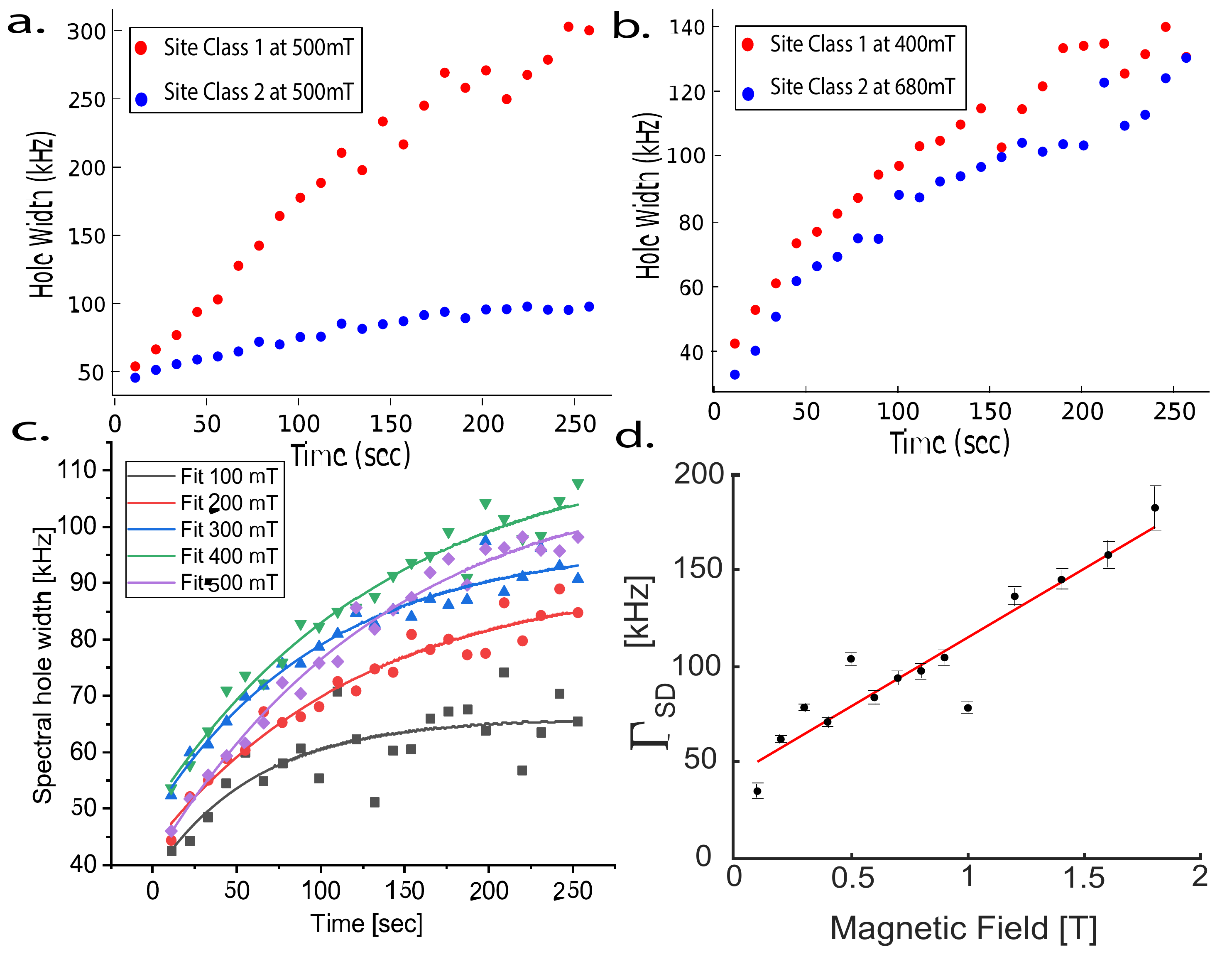}
\caption { \textbf{a.} Spectral hole widths for ions in both magnetic classes as a function of time at 500mT. The class of ions with the larger magnetic field projection shows significantly more spectral diffusion. \textbf{b.} Spectral hole widths for ions in both magnetic classes as a function of time where the applied magnetic field for each class of ions creates roughly equivalent splitting. The spectral diffusion is comparable. \textbf{c.} Fitted Hole width versus time for a single class of ions at a series of different magnetic fields. \textbf{d.} Linear fit to the variation of $\Gamma_{SD}$ as  a function of magnetic field.}
\label{spectraldifffig}
\end{centering}
\end{figure}

     The first portion of evidence for determining the source of the measured diffusion is the different behavior of each magnetic class of ions. Shown in Fig. \ref{spectraldifffig} a is the hole width of each such class as a function of time for an applied field of 500mT. For the same external field each class broadens at a different rate. However, when different fields are applied, such that the Zeeman splittings of each class are roughly equivalent, the broadening occurs at the same rate, see Fig. \ref{spectraldifffig} b. This contrast likely rules out contributions of two level systems \cite{Thiel_2014} on this time scale, which are likely to be located randomly throughout the material without an orientation dependent response. Instead, the diffusion appears to depend upon the magnitude of the created magnetic dipole moment of the Tm ions.  
    
    To further isolate the cause of the diffusion we analyze the behavior of a single class of ions as a function of applied field. Utilizing the model in Eq. \ref{sudden Jump} we fit $\Gamma_{SD}$ and $R_s$ exclusively for each magnetic class of ions where the functional form of $\Gamma_{SD}$ is given by \cite{Bottger2006},
\begin{align} \label{sech}
\Gamma_{SD}(B,T)= \Gamma_{max}\operatorname{sech}^2(\frac{g_{env}\mu_{B}B}{kT})
\end{align}
    Here  $g_{env}$ is the g-factor of spins in the local crystalline environment, $\mu_{B}$ is the Bohr magneton, B is the applied field, k is the Boltzmann constant, T is the sample temperature, and $\Gamma_{max}(B)$ is the full width half max of the broadening of the optical transition due to external sources. 
    
    For non-Kramer's ions in general, and all atomic species in the YGG host, since the g-factors, $g_{env}$, are quite small, the thermal distribution of population in various spin levels remains nearly constant for the temperatures and applied fields used in this work, making the $\operatorname{sech}^2$ factor in Eq.\ref{sech} effectively constant as well. Any field dependence of $\Gamma_{SD}$ can then be attributed to the field dependence of $\Gamma_{max}(B)$. As shown in Fig. \ref{spectraldifffig} d we find $\Gamma_{SD}$ to increase as the magnetic field increases while $R_s$ remains fairly constant for the measured fields. Inspired by \cite{Bottger2006,Lutz2016} we fit this behavior using a functional form of $\Gamma_{max}(B)$ that relates the optical transition broadening to the quadratic Zeeman effect.
\begin{align}
\Gamma_{max}(B) = g_j^2 \mu_{B}^2 B \cdot |\boldsymbol{\Lambda_e}-\boldsymbol{\Lambda_g}| \cdot B_{Noise} + \Gamma_{max}(0).
\end{align}
    Here, $g_j$ is the electronic g factor for each level, $\mu_{B}$ the Bohr magneton, $\boldsymbol{\Lambda}_i, \ i \in \{g,e\}$ is the hyperfine tensor for the ground and excited states in the transition, $B$ is the applied magnetic field, and $B_{Noise}$ the width of the field distribution around each experimentally set value due to flips of local host spins or external noise.  All factors in this expression are known save this empirical $B_{Noise}$, which is fit to a value of 87$\pm$20~$\mu$T using the data of Fig. \ref{spectraldifffig} d. 
    
\begin{center}
\begin{table*}
\centering
\begin{tabular}{|c|c|c|c|c|} 
\hline
 Species & Concentration & Effective g factor & Average distance (r)  &Average Field (B)  \\
\hline
$^{169}$Tm$^{3+}$ & 1\% & 0.0077 & 17 \r{A}  & 500 nT \\ 
$^{71}$Ga & 40\% & 0.00071  & 2.6 \r{A} & 15 $\mu$T \\   
$^{69}$Ga & 60\% & 0.00092 & 2.6 \r{A}  & 19 $\mu$T \\   
$^{89}$Y & 99\% & 0.00014 & 4 \r{A}  & 850 nT\\ 					
\hline
\end{tabular}
\caption{ Magnetic fields experienced at an average Tm ion site due to spin flips of the elemental species in the crystal. Estimated distances are in agreement with measured data from YAG \cite{euler1966oxygen}.  Only the gallium spins match the order of magnitude of the measured magnetic variations.} 
 \label{FieldTable}
\end{table*}
\end{center}
    
%\begin{center}
%\begin{table*}
%\centering
%\begin{tabular}{|c|c|c|c|c|c|} 
%\hline
% Species & Concentration & Effective g factor & Dist. A &  Dist. B & Dist. C \\
%\hline
%$^{169}$Tm$^{3+}$ & 1\% & 0.0077 & 17.47 \r{A}  & - & 3.16 \r{A}  \\ 
%$^{71}$Ga & 40\% & 0.00071  & 2,88 \r{A} &  2.99 \r{A}& 1.48 \r{A} \\   
%$^{69}$Ga & 60\% & 0.00092 & 2.88 \r{A}  &  2.99 \r{A} & 1.48 \r{A}\\   
%$^{89}$Y & 99\% & 0.00014 & 3.92 \r{A}  &  3.67\r{A} & 0.84 \r{A}\\ 					
%\hline
%\end{tabular}
%\caption{\textcolor{red}{Anta/Nir which of these tables and captions do you like?} Estimates of the relative distance between various crystal spins and an average Tm ion site based on calculations from \cite{Bottger2002}, a model of the YGG crystal cell \cite{Persson20142}, and measured bond lengths in YAG \cite{Giorgetti2009}. Column \textbf{Dist. A} estimates distance from an average Tm ion site to each spin species based on relative concentrations and unit cell dimensions. Column \textbf{Dist. B} are the literature values from YAG garnet crystals for the various crystal distances. Column \textbf{Dist. C} computes the average distance using g-factors for spin flips of each host crystal constituent, and the crystal model, to estimate the expected distance to produce 87uT of field noise from spin flips near the Tm ion site. The three estimation methods are only in agreement for the gallium spins in Tm:YGG.} 
% \label{FieldTable2}
%\end{table*}
%\end{center}
    Following calculations from \cite{Bottger2002} and using a model of the YGG crystal cell from \cite{Persson20142} we estimate the magnetic field variation at an average Tm ion site given the concentrations, unit cell dimensions, and corresponding g-factors of each host crystal constituent. Table \ref{FieldTable} contains estimations of magnetic fields at a potential Tm ion site due to spin flips from each of the crystal constituents. Considering possible contributors, the low Tm ion concentration of this crystal means the average distance to another dopant thulium is likely too small for Tm-Tm interactions to cause noise of this magnitude. Yttrium, though fully concentrated, is also too weakly magnetic. The most likely explanation is the presence of a fully concentrated bath of local gallium spins, each with a moderate nuclear magnetic moment. An additional contribution from a noisy current supply used to power the superconducting solenoid that creates the magnetic field across the crystal, among other factors stemming from geometrical concerns and heterogeneous spin species \cite{van1948dipolar}, may explain the mismatch between our rough estimations of spin flip noise, and the fitted field variations.

\section{Conclusion}
	
We conducted a series of spectroscopic measurements that lead to a detailed understanding of how to create high-resolution spectral features using Tm:YGG. The possibility for controlling pump field amplitude, frequency chirp, pulse duration, and pulse power give a number of experimental handles with which to optimize the shape of the desired spectral feature. For zero magnetic field experimental results were shown to be in good agreement with predictions from a three level, frequency dependent, rate equation model for various sets of input pulses shorter than the transition lifetime. However, when applying a similar model to a three level system that includes a ground state spin level split by magnetic field, magnetic noise, and spectral diffusion dominate the resulting features over long timescales. We attribute large shifts to each ions optical transition frequency due to quadratic Zeeman effect in combination with time varying magnetic fields as the likely cause of the diffusion. This in depth analysis will aid applications that employ spectral features created using Tm:YGG and other rare earth doped crystals such as spectral hole-based laser stabilization, RF frequency analysis, and optical quantum memories. 
	
\section*{Acknowledgements}
 This material is based in part on research at Montana State University sponsored by Air Force Research Laboratory under Agreement No. FA8750-20-1-1004. 

%

%\bibliography{Bib2.bib}

\section{Supplementary Material} \label{Supplement}

\subsection{More examples of burning sequences}
    In the main text we show examples of modeled and experimentally created spectral holes using a single burning process. However, the good agreement between these two methods extends to a large variety of procedures. We measured holes created using burn pulses with 5 different optical powers, with 2 different durations, and 4 different waiting times, giving us more than 40 individual experiments, and innumerable data points using which we can match behavior and predictions. 
%Results of
%\begin{figure}[h!]
%\begin{centering}
%\includegraphics[width=\textwidth]{chapter-7/SI_More_Surfaces.pdf}
%\caption {\textbf{a.} }
%\label{blah}
%\end{centering}
%\end{figure}

\subsection{Holes relative to Level Lifetimes}
    All the cases listed in the previous section showcase the behavior of spectral holes relative to the different level lifetimes. They are worth a deeper look. We measure only a pair of burning times, 100us and 1ms, as burning pulses longer than around 1 ms were observed to heat the crystal. For burning pulses of each length we examine a set of different waiting times between the pulses of the sequence shown in Fig \ref{HoleSurface} a : 1ms, 10ms, 50ms, and 100ms. Each is selected for a reason. One millisecond is on the order of the the excited state lifetime and much shorter than the bottleneck lifetime. Ten milliseconds is much longer than the excited lifetime and much shorter than the $^3$F$_4$ bottleneck lifetime. Fifty milliseconds is much longer than the excited and on the same order as the bottleneck lifetime, and 100ms is much longer than both lifetimes.
    
\begin{figure}[h]
\begin{centering}
\includegraphics[width=0.5\textwidth]{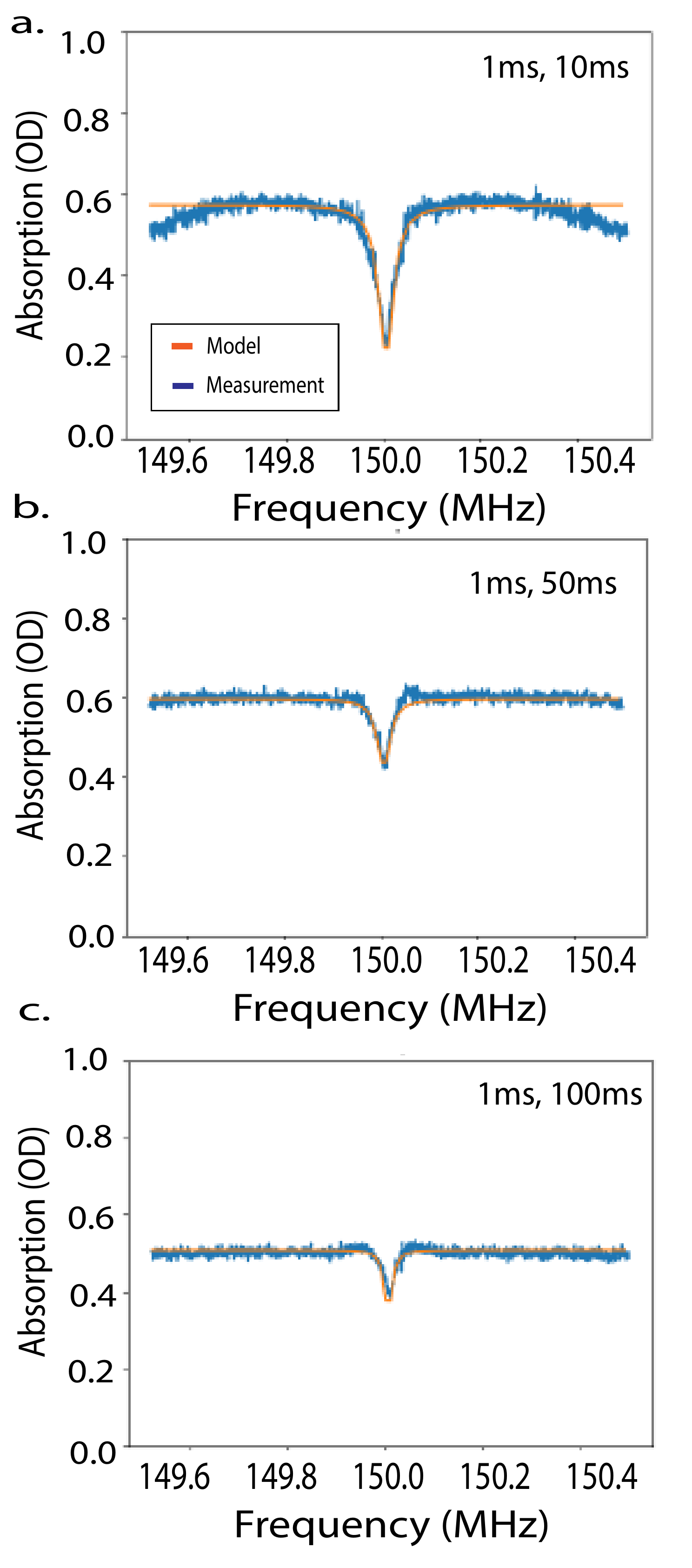}
\caption { Spectral holes after 50 cycles of 1 ms burning separated by \textbf{a} 10 ms waits, \textbf{b} 50 ms waits, \textbf{c} 100 ms waits. The hole decreases in total depth because population decays from the bottleneck level before the next pumping step.}
\label{DifWaits}
\end{centering}
\end{figure}

    As the waiting time approaches the bottleneck level lifetime the additive nature of multiple burn pulses start to disappear and the hole depth reaches a much smaller steady state. This can be seen from Fig. \ref{DifWaits} a-c. When the waiting time exceeds the bottleneck lifetime we are essentially re-pumping the same population with each burn pulse, interacting with a nearly unchanged material absorption profile. For waiting times on the order of the excited state lifetime, we see a slight broadening of the hole due to power broadening as well as frequency shifts caused by excited ions within the pump bandwidth \cite{Rippe2005,Roos2004}. This indicates that the optimum wait duration using which to create features is longer than the excited state lifetime so that most pumping occurs without population in the excited state, but still much shorter than the bottleneck level lifetime to allow population to accumulate in the bottleneck over many successive burning pulses.

\subsection{Power Dependence Due to Model Non-Linearity}

\begin{figure}[h!]
\begin{centering}
\includegraphics[width=0.5\textwidth]{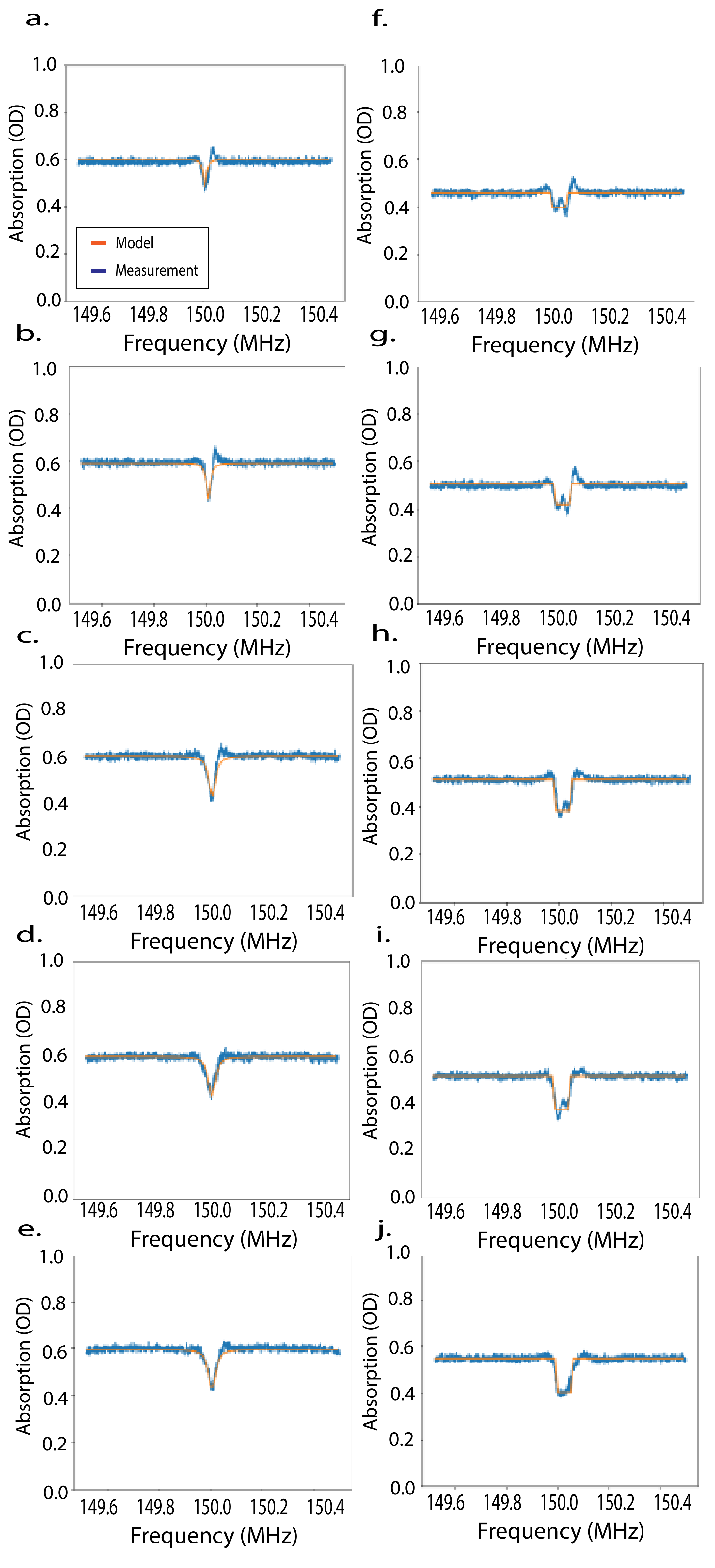}
\caption {\textbf{a-e.} A series of spectral holes after 50 cycles of 1 ms burning separated by 50ms waiting between for five linearly increasing optical powers. Points(blue) are measured spectral holes, with line (orange) results of rate equation model for this sequence and the pumping line-shape.  \textbf{f-j.} A similar experiment, but each burn pulse is now a hyperbolic secant pulse.}
\label{powBroad}
\end{centering}
\end{figure}

    Of the many measured data sets, a key portion involves the same optical pumping sequence with a series of different optical powers, See Fig. \ref{powBroad}. This tuning is done by changing the peak to peak amplitude of the AOM driving signal over a range of voltages. Different initial laser line-shapes, $R(\Delta)$ with different amplitudes broaden the resulting holes at different rates. This study allowed us to determine the spectral shape of our applied $R(\Delta)$ function by comparing measured results to the scaling predicted by our model. First, in Fig. \ref{powBroad} a-e. we show a series of hole burning experiments conducted with different optical powers using our un-modulated laser. The clear broadening of the spectral hole is not  power broadening as in \cite{Allen2012} but rather an effect due to the Lorentzian shape of the applied laser line. With increased amplitude the wings of the Lorentzian shape gain the ability to efficiently change ground state population, broadening the resulting hole. 
    Additionally, Fig. \ref{powBroad} b. shows the scaling of a spectral hole for hyperbolic secant modulated pulses using a burning sequence with matching timing as above. As seen in figure \ref{powBroad} f-j, the broadening is nearly absent because the pulse modulation reshapes the applied $R(\Delta)$ function into a square shape that does not possess wings that strengthen with added burning power. This is the more desirable scaling, so that added power results in a higher excitation rate and deeper rather than broader spectral holes.

\subsection{ Expected Quadratic Zeeman Shifting Behavior}
First, we can calculate the initial frequency of in-homogeneous line for some initial field magnitude and orientation. This is given by the difference between the shift of the ground ($D_g$) and excited($D_e$) states. 
\begin{align}
    D_J= \frac{g_j \mu_B}{2A_J}[ (\gamma_{J,x}- \gamma_n)B_x^{2}+ (\gamma_{J,y}- \gamma_n)B_y^{2} + (\gamma_{J,z}- \gamma_n)B_z^{2}]
\end{align}
Here, $g_n$ is the nuclear gyro-magnetic ratio of thulium,
$\beta_n$ is the nuclear magneton, $g_J$ is electronic g factor for
each level, $\mu_B$ is the Bohr magneton, and $A_J$ is the hyperfine
interaction constant, and $B_{(x,y,z)}$ the field components in the local frame of the ions. 

Then the initial hole is at a frequency $\Delta_i= \frac{D_{ei}-D_{gi}}{\hbar} $
Calculating a $\Delta_f= \frac{D_{ef}-D_{gf}}{\hbar} $ using the final  field magnitude and orientation we arrive at the projected shift $\Delta_f-\Delta_i$ in hole center frequency due to the quadratic Zeeman effect.

\end{document}